\documentclass[prl,showpacs,prl,superscriptaddress,twocolumn]{revtex4}

\usepackage{amsfonts}
\usepackage{amsmath}
\usepackage{amssymb}
\usepackage{graphicx}

\usepackage{color}

\begin{document}

\title{Acoustic attenuation probe for fermion superfluidity in
ultracold atom gases}

\author{Sergio~Gaudio}
\thanks{Now at Universita' degli Studi di Roma, La Sapienza, and  Institute of Complex Systems, CNR, Roma, 00185, Italy\\
Electronic address: gaudios\protect{@}roma1.infn.it
}
\affiliation{Department of Physics,
     Boston College,
     Chestnut Hill, MA 02167}
\affiliation{Theoretical Division,
     Los Alamos National Laboratory,
     Los Alamos, NM 87545}

\author{Bogdan Mihaila}
\affiliation{Theoretical Division,
     Los Alamos National Laboratory,
     Los Alamos, NM 87545}

\author{Krastan~B.~Blagoev}
\affiliation{Theoretical Division,
     Los Alamos National Laboratory,
     Los Alamos, NM 87545}
\author{Kevin~S.~Bedell}
\affiliation{Department of Physics,
     Boston College,
     Chestnut Hill, MA 02167}
\author{Eddy Timmermans}
\thanks{Electronic address: eddy\protect{@}lanl.gov}
\affiliation{Theoretical Division,
     Los Alamos National Laboratory,
     Los Alamos, NM 87545}
\begin{abstract}
Dilute gas Bose-Einstein condensates (BEC's), currently used to cool
fermionic atoms in atom traps, can also probe the superfluidity of
these fermions. The damping rate of BEC-acoustic excitations (phonon
modes), measured in the middle of the trap as a function of the
phonon momentum, yields an unambiguous signature of BCS-like
superfluidity, provides a measurement of the superfluid gap
parameter and gives an estimate of the size of the Cooper-pairs in
the BEC-BCS crossover regime. We also predict kinks in the momentum
dependence of the damping rate which can reveal detailed information
about the fermion quasi-particle dispersion relation.
\end{abstract}
\maketitle
Recently, cold atom physics entered~\cite{experiments1} the
previously inaccessible~\cite{Eagles2} Bose-Einstein condensate
(BEC) regime of fermion superfluidity. A magnetic field sweep through a Feshbach
resonance induced the fermions to pair up into diatomic molecules
which subsequently occupied a BEC (i.e. forming pairs
hardly distinguishable from diatomic molecules -- Shafroth pairs).
Current experiments aim at taking the fermion superfluid into the
Bardeen-Cooper-Schrieffer (BCS) limit of ordinary superconductors,
in which the fermion pairs are larger than the average interparticle
distance (Cooper pairs).  The crossover region~\cite{crossover},
near the detuning of diverging scattering length, is of fundamental
interest~\cite{comment}.  The cold atom crossover
experiments~\cite{experiments} illustrate the need for unambiguous
probes of fermion superfluidity.
Sympathetic cooling immerses the fermionic atoms in colder
bosonic atoms, often occupying a BEC.  The system then resembles the
condensed $^{3}$He-$^{4}$He liquid mixtures~\cite{Saarela}.  In
contrast to the helium liquid mixtures in which strong interaction
effects drastically reduce the boson-mediated fermion-fermion
interactions~\cite{Bardeen}, the cold atom mixture can exhibit
BEC-mediated fermion superfluidity at attainable
temperatures~\cite{Heiselberg}. 

\begin{figure}[t]
   \includegraphics[width=\columnwidth]{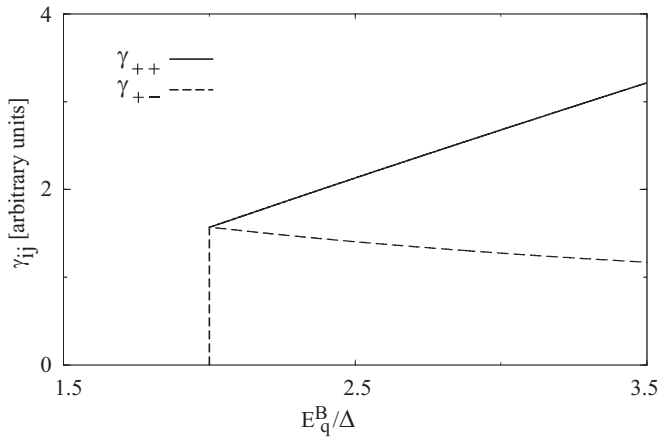}
   \caption{\label{fig:scattering_lengths}
Scaled damping rates components, $\gamma_{++}$ and $\gamma_{--}$, in
the BCS limit, as a function of the phonon energy, $E^{B}_{q}$,
measured in units of $\Delta$, the superfluid gap parameter. Here,
we have $\gamma_{ij}=\Gamma_{ij}\sqrt{1+(\xi q)^{2}}/\Gamma_{0}$.}
\end{figure}

In this paper, we show that the BEC simultaneously trapped with the
fermion atoms, can provide a powerful probe of fermion
superfluidity.  We study the damping rate, $\Gamma(q)$, of acoustic
BEC-excitations (phonons) in the mixture as a function of the phonon
momentum $q$.  Acoustic attenuation~\cite{acoustic}
provides a signal for fermion superfluidity, determines the
superfluid gap parameter, tests the validity of the weakly
interacting quasi-particle description~\cite{Kevin}, estimates the size of the
Cooper-pairs and probes the quasi-particle dispersion relation. The
latter could be inferred from
kinks which we predict to appear as an imprint of kinematical
constraints.\\
In linear response theory $\Gamma(q)$ is related to density-density
correlations. For a sufficiently large trap, the damping rate
measured near the middle of the trap, where the particle densities
are approximately constant, equals that of a BEC-phonon in a
homogeneous mixture.
The BEC-phonon is a Bogoliubov excitation with energy  $E^{B}_{q} =
\hbar q c \sqrt{1+ (\xi q)^{2}}$, where $c$ denotes the BEC sound
velocity, $c=\hbar/(2m_{B}\xi)$, $\xi$ is the BEC-healing length,
$\xi = (16 \pi \rho_{B} a_{B} )^{-1/2}$, and $a_{B}$ is the
boson-boson scattering length. Using box normalization, the boson
density $\rho_{B}$ corresponds to $N_{B}$ bosons condensed in a
volume $\Omega$, $\rho_{B}=N_{B}/\Omega$. The system also includes a
homogeneous mixture of spin $+$ and $-$ fermions, with
$\rho_{+}=\rho_{-}=\rho_{F}$.
Each single-component fermion system has Fermi wavenumber $k_{F} =
(6 \pi^{2} \rho_{F})^{1/3}$, Fermi-velocity $v_{F}=\hbar
k_{F}/m_{F}$ and Fermi energy $\epsilon_{F} = \hbar^{2}
k_{F}^{2}/(2m_{F})$.  The boson-fermion interaction is spin
dependent, giving different boson-fermion scattering lengths,
$a_{+}$ and $a_{-}$, and a fermion-boson interaction Hamiltonian
$\hat{H}_{int} = (2 \pi \hbar^{2} \overline{a}) (1/m_{B} +1/m_{F})
\Omega^{-1} \sum_{{\bf k},j} \hat{\rho}^{B}_{-{\bf k}} \alpha_{j}
\hat{\rho}^{F}_{j,{\bf k}}$, where $j=\pm$, $\hat{\rho}^{B(F)}_{{\bf
k}}$ represents the boson (fermion) density operator, $\overline{a}
= \sqrt{a_{-}^{2}+a_{+}^{2}}$, and $\alpha_{j} =
\overline{a}/a_{j}$. We use perturbation theory around
noninteracting fermion-boson product states, $|\rangle_{F} \otimes
|\rangle_{B}$.  The $\hat{H}_{int}$--interaction then annihilates a
BEC-phonon $|{\bf q}\rangle_{B}$ and creates a fermion excitation
that is a collective mode or a quasi-particle pair of opposite
spins.  By ignoring initial states with multiple BEC excitations, we
assume sufficiently low temperatures.  By ignoring final states with
two (or more) phonons, we make a relative error $\propto (q\xi)^{3}
(a_{B}/\xi) (8/3\pi)$, but we take $(q\xi)< (\xi/a_{B})^{1/3}$.
Using $_{B}\langle 0| \hat{\rho}_{-{\bf q}} | {\bf q} \rangle_{B} =
\sqrt{N_{B} (\hbar^{2} q^{2}/2 m_{B})/(\hbar \omega^{B}_{q})}$,
$\Gamma(q)$ takes the form $\Gamma({\bf q}) = \sum_{i,j} \alpha_{i}
\alpha_{j} \Gamma_{ij}({\bf q})$ with

\begin{align}
\Gamma_{ij} ({\bf q}) = & \ \frac{8}{3}\frac{ (\Gamma_{0}/4\pi) \xi
q}{\sqrt{1+(\xi q)^{2}}}
\int \frac{d^{3} r}{\rho_{F}}  \int \frac{dt}{\hbar/\epsilon_{F}}
\label{gamma1}
\\ \nonumber & \times \
\exp\bigl [-i({\bf q} \cdot {\bf r} - \omega^{B}_{q} t)\bigr ] \
\langle \hat{\rho}^{F}_{i}({\bf r},t) \hat{\rho}^{F}_{j}({\bf 0},0) \rangle \;
\end{align}
where  $\Gamma_{0} = 4 \pi \overline{a}^{2} \rho_{B} v_{F}
[(1+m_{F}/m_{B})/2]^{2}$, proportional to the rate at which a
particle of velocity $v_{F}$ scatters off hard-core targets of
radius $\overline{a}$, distributed at the BEC density.  By varying
$a_{+}$ or $a_{-}$, one can, in principle, distinguish the
$\Gamma_{++}$ and $\Gamma_{+-}$ contributions. In a Cooper-paired
fluid, the $\langle \hat{\rho}^{F}_{+} \hat{\rho}^{F}_{-} \rangle$
correlations depend on the microscopic superfluid structure,
embodied, for instance, by the order parameter (i.e. the expectation
value of annihilation field pairs,  $\langle \hat{\psi}_{+}
\hat{\psi}_{-} \rangle \neq 0$). The equal time $\pm$-density
fluctuation, $\langle \hat{\rho}^{F}_{+}({\bf r})
\hat{\rho}^{F}_{-}({\bf 0}) \rangle \approx \langle
\hat{\psi}_{+}({\bf r}) \hat{\psi}({\bf 0}) \rangle \langle
\hat{\psi}^{\dagger}_{-}({\bf r}) \hat{\psi}^{\dagger}({\bf 0})
\rangle = F({\bf r}) F^{\star}(-{\bf r})$, where $F({\bf r}) =
\langle \hat{\psi}_{+}({\bf r}) \hat{\psi}_{-}({\bf 0}) \rangle$
plays the role of fermion pair wavefunction.  Equation
(\ref{gamma1}) also suggests that one can probe directional
information if and when cold atom experiments realize anisotropic
fermion superfluidity (such as p-wave, d-wave or
LOFF-pairing~\cite{fulde,Wilczek}). To make the quasi-particle pair
explicit, we insert a summation over such states in the
density-density correlation function.   The temporal integral gives
a delta function that expresses the kinematics of the pair creation
process,
\begin{align}
\label{gamma}
& \Gamma_{ij} ({\bf q})
\\ \notag & \! =
\frac{(8\Gamma_{0}/3)\xi q}{\sqrt{1+(\xi q)^{2}}} \,
\frac{\epsilon_{F}}{\rho_{F}} \!
\int \!\!\frac{d^{3}k}{(2\pi)^{3}} \,
g_{i,j}({\bf k};{\bf q}) \,
\delta \bigl ( E^{B}_{q} \!-\! E^{F}_{k} \!-\! E^{F}_{|{\bf k}-{\bf q}|}
\bigr )
\end{align}
where $E^{F}_{k}$ denotes the quasi-particle dispersion relation and
$g_{i,j}({\bf k};{\bf q})$ is the weight for the creation of
quasi-particles of momenta ${\bf k}$ and ${\bf q}-{\bf k}$. The
generalized Wick's theorem leads to $g_{++}=g_{--}= n_{{\bf k}}
(1-n_{{\bf q}-{\bf k}})$, and $g_{+-}=g_{-+}= F_{{\bf k}}
F^{\ast}_{{\bf q}-{\bf k}}$, with $F({\bf r}) = (2\pi)^{-3} \int
d^{3} k F_{{\bf k}} \exp(i {\bf k} \cdot {\bf r})$, and $n_{{\bf
k}}=\langle {c}^{\dagger}_{j,{\bf k}} {c}_{j,{\bf k}}\rangle$. Here,
$c (c^{\dagger})$ denote fermion annihilation (creation) operators.\\
The form of Eq.~(\ref{gamma}) is quite general but the lack of
correlations resulting from Wick's rule, is not.  We can understand
the role of $F$ as pair wavefunction by comparing Eq.~(\ref{gamma})
to the Fermi Golden-rule rate at which a particle of momentum ${\bf
q}$ and energy $E^{B}_{q}$ scatters to zero momentum by breaking the
bond of stationary diatomic molecules of $(+)$ and $(-)$ atoms. In
the initial molecules, these atoms occupy bound vibrational states
with wavefunction $\varphi_{{\bf k}}$ and binding energy $E_{b}$.
The interaction that breaks the molecular bond, $\hat{H}_{int}$,
creates a pair of free atoms, the $(+)$ atom  with momentum ${\bf
k}$ and energy $e_{k}$, and the $(-)$ atom with momentum ${\bf
q}-{\bf k}$ and energy $e_{|{\bf q}-{\bf k}|}$. The corresponding
interaction matrix element is proportional to $\alpha_{-}
\varphi_{{\bf k}} + \alpha_{+} \; \varphi_{{\bf k}-{\bf q}}$, so
that the size-dependent contribution to the rate is
\begin{align}
\frac{1}{\tau_{q}} \propto \alpha_{+} & \alpha_{-} \!\! \int d^{3}k
\, \varphi_{{\bf k}} \varphi_{{\bf k}-{\bf q}}^{\ast} \delta (E_{q}
\!- \! E_{b} - e_{k} - e_{|{\bf k}-{\bf q}|} ) \>, \label{bb}
\end{align}
which has the form of Eq.~(\ref{gamma}) if we associate $F_{{\bf k}}
\rightarrow \varphi_{{\bf k}}$ and $E^{F}_{{\bf k}} \rightarrow
E_{b}/2 + e_{k}$.  Similar to the $g_{+-}$ contribution to
Eq.~(\ref{gamma}), Eq.~(\ref{bb}) stems from an interference of two
processes in reaching the final state.  In the first process, atom
$(+)$ with momentum ${\bf k}$ and atom $(-)$ with momentum $-{\bf
k}$ in the molecule reach the final state because the $(-)$ atom
receives a momentum kick ${\bf q}$.  In the second process, the
$(+)$ and $(-)$ atoms in the molecule have initially momenta ${\bf
k}-{\bf q}$ and ${\bf q}-{\bf k}$, respectively, and the $(+)$ atom
receives a momentum kick~${\bf q}$.\\
The role of the fermion density of states, $D(E) = (2\pi^{2})^{-1}
k^{2}(E)/|\partial E^{F}/\partial k|$, becomes apparent when we
convert Eq.~(\ref{gamma}) to an integral over quasi-particle energy.
Targeting s-wave pairing, we assume a spherically symmetric pairing
wavefunction, $F_{{\bf q}} = F_{q}$.  Introducing the angle $\theta$
between ${\mathbf k}$ and ${\bf q}$, we write
\begin{eqnarray}
\delta\left[
E^{B}_{q} - E^{F}(k) - E^{F}\Bigl(
\sqrt{k^{2}+q^{2}-2kq\cos\theta}\Bigr) \right]
\nonumber \\
   =
\sum_{j} \frac{|{\bf q}-{\bf k}|_{j} \
\delta(\cos\theta-\cos\theta_{j})}{k_{j} q
\left| \partial E(\kappa)/\partial \kappa \right|_{\kappa=|{\bf
q}-{\bf k}|_{j}} } \; ,
\label{der}
\end{eqnarray}
where the $j$-subscript enumerates the energy and momentum
conserving quasi-particle pairs.  Whether or not the fermion
dispersion relation supports a local minimum (i.e., whether or not
the fermion superfluid resides in the BCS regime), affects the
number of such pairs.  If $E^{F}_{k}$ has a local minimum, we refer
to it as the ``gap''~$\Delta$, $E^{F}_{k_{m}}=\Delta$ and we refer
to the corresponding surface in momentum space, $k=k_{m}$, as the
gap surface.  We can then assign two momentum magnitudes for the
same quasi-particle energy $E$: $k_{<}(E)$ inside the gap surface
($k_{<}(E)<k_{m}$) and $k_{>}(E)$ outside ($k_{>}(E)>k_{m}$).  The
phonon annihilation can generate four quasi-particle pair types,
$j=(<,<),(<,>),(>,<),(>,>)$, where we indicate the momentum position
of the (+) particle first, and that of the (-) particle second.
Depending on the energy distribution, some pair creation processes
$j$ are excluded kinematically.  We can picture the constraints
geometrically for the j-pair creation by considering the spherical
surfaces in momentum space of radii $k_{j,+}(E)$ and $k_{j,-}(E')$,
with $E+E'=E^{B}_{q}$. As the $(+)$ momentum ${\bf k}$ lies on the
first surface, the $(-)$ momentum ${\bf q}-{\bf k}$ on the second,
the surfaces must be connected by the ${\bf q}$-momentum, requiring
$k_{j,+}(E)+k_{j,{-}}(E') < q <|k_{j,+}(E)-k_{j,-}(E')|$.  Then, the
damping rates read
\begin{align}
\frac{\gamma{+-}(q)}{\left(  \frac{8 \pi^{2} \xi \epsilon_{F}}{3
\rho_{F}} \right)} =  & \sum_{j} \!     \int \! dE \, \frac{D_{j}(E)
F_{j}(E)}{k_{j}(E)} \frac{D_{j}(E') F_{j}(E')}{k_{j}(E')} \>,
\label{frk}
\\ \notag
\frac{\gamma_{++}(q)}{\left(  \frac{8 \pi^{2} \xi \epsilon_{F}}{3
\rho_{F}} \right)} = & \sum_{j} \! \int \! dE \, \frac{D_{j}(E)
n_{j}(E)}{k_{j}(E)} \frac{D_{j}(E') \left[1 \!- \! n_{j}(E') \right]
}{k_{j}(E')} ,
\end{align}
where 
$\gamma_{ij}(q)=[\Gamma_{ij}(q)  \sqrt{1+(\xi q)^{2}}]/\Gamma_{0}$.
What is the information content of $\Gamma(q)$?  The gap prevents
soft phonons with $E^{B}_{q}<2\Delta$ from damping through pair
excitation.  If these soft phonons are kinematically prohibited from
exciting collective fermion modes because the latter's velocity
($\sim v_{F}$ for the Anderson-Goldstone mode in the BCS limit)
exceeds $c$, then they can only decay through Beliaev damping which
gives a negligible damping rate at long wavelengths ($\Gamma_{\rm
Beliaev}(q) \sim (6\sqrt{\pi}/5)
\sqrt{\rho_{B}a_{B}^{3}}(qc)(q\xi)^{4}$). More energetic BEC phonons
damp by radiating quasi-particle pairs, and the higher the density
of states, the more an energy interval contributes to the damping.
At the gap surface the density of states is infinite, although the
divergence is sufficiently slow ($\propto \sqrt{E^{2}-\Delta^{2}}$)
to ensure that the integral~(\ref{frk}) remains finite.  The
divergence plays a 
prominent role when both quasi-particles form on or near the gap
surface, corresponding to a phonon energy that slightly exceeds $2
\Delta$.  While the damping rate remains finite, the diverging
density of states causes a discontinuous jump of the damping rate at
the (edge) phonon momentum $q_{\Delta}$ for which
$E^{B}_{q_{\Delta}}=2\Delta$,
see Fig.~\ref{fig:scattering_lengths}.
Conversely, the discontinuity in the damping rate points towards a
diverging density of states, providing a signature of fermion
superfluidity on the BCS side of the crossover.  This sharp feature
also provides a sensitive way of determining the superfluid gap from
the edge momentum $q_{\Delta}$, $\Delta=(\lambda_{B} \rho_{B}) (\xi
q_{\Delta}) \sqrt{1+(q_{\Delta}\xi)^{2}}$. On the other hand, the
observation of a continuous onset of the damping with increasing $q$
does {\it not} necessarily imply that the fermions, if superfluid,
have entered the BEC-side of the crossover.  Indeed, as the system
nears the crossover from the BCS side in an adiabatic sweep through
a Feshbach resonance, we expect $k_{m}$ to continuously tend to
zero, whereas $q_{\Delta}$ monotonically increases.  Beyond the
detuning for which $k_{m} = q_{\Delta}$, $q_{\Delta}$ is too large
to fit inside the gap surface.  In that case, the creation of both
quasi-particles on the gap surface, the event that leads to the
discontinuous jump in damping rate, is kinematically prohibited.
\begin{figure}[b]
   \includegraphics[width=\columnwidth]{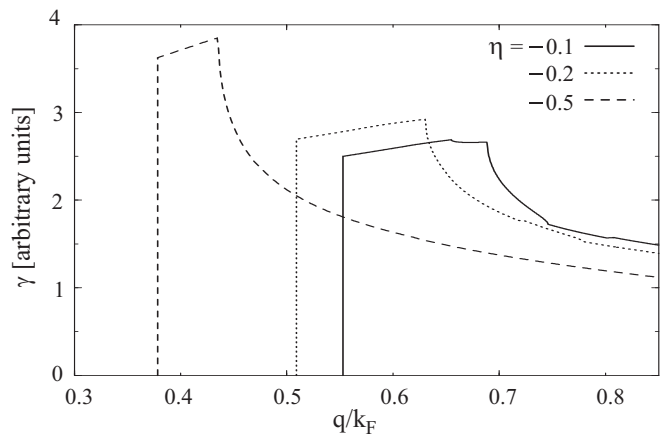}
   \caption{\label{fig:DampingRates}
Near-edge pair breaking contribution to the scaled damping rate
$\gamma$ ($\gamma=\Gamma \sqrt{1+(\xi q)^{2}}/\Gamma_{0}$),  as a
function of the phonon momentum, $q$. Here we assumed $a_{+}=a_{-}$,
$c/v_{F}=0.5$ and varied the fermion-fermion scattering lengths
$a_{F}$ or $\eta=(k_{F}a_{F})^{-1}$. The edge momentum gives the
superfluid gap, the slope near the edge relates to the size of the
Cooper-pairs and the discontinuities in the derivatives (kinks) are
imprints of the kinematical constraints of the pair breaking
process. }
\end{figure}
The dependence of the damping rate on Cooper-pair size, suggested by
the analogy with breaking the stationary molecule bond,
Eq.~(\ref{bb}), becomes apparent when we expand the damping rates
near the edge.  We make two assumptions: (i) $\left(E^{F}_{k}
\right)^{2} \approx \Delta^{2} + \hbar^{2} v^{2} (k-k_{m})^{2}$, in
which $v$ characterizes the curvature of the dispersion near the
gap.  At $k=k_{m}$, $F(k_{m})= F_{m}$ and $n(k_{m})=n_{m}$. (ii) The
fermion system is sufficiently far from the crossover point to
ensure that all four quasi-particle pair events are kinematically
allowed.  Therefore, the damping rates near $q=q_{\Delta}$,
($q>q_{\Delta}$) are
\begin{align}
& \frac{\gamma_{+-}(q)}
{2\pi^{2} (\xi k_{F}) (\Delta/[m_{F}v^{2}/2])}
\frac{1}{F_{m}^{2}}
\label{edge+-}
\\ \nonumber
& \quad \approx
   1 + \frac{E^{B}_{q}-2\Delta}{\Delta}
\biggl [ \frac{3}{4} + \frac{\Delta^2}{(\hbar v)^2} \frac{
\nabla^{2}_{k} F_{k} }{F_{k}} \Bigr |_{k=k_{m}} \biggr ] \>, \\
&
\frac{\gamma_{++}(q)} {2\pi^{2} (\xi k_{F}) (\Delta/[m_{F}v^{2}/2])}
\frac{1}{n_{m} (1-n_{m})} \label{edge++}
\\ \nonumber
& \quad \approx
1 + \frac{E^{B}_{q}-2\Delta}{\Delta}
\biggl [ \frac{3}{4} + \frac{\Delta^2}{(\hbar v)^2}
\frac{1-2n_{m}}{1-n_{m}}
\frac{ \nabla^{2}_{k} n_{k} }{n_{k}} \Bigr |_{k=k_{m}} \biggr ]
\>.
\end{align}
Note that the ratio of the edge values of the $(+,+)$ and $(+,-)$
damping rates connect $n_{m}$ and $F_{m}$:
   $\Gamma_{+-}(q_{\Delta})/\Gamma_{++}(q_{\Delta}) =
n_{m} (1-n_{m})/F_{m}^{2}$.  This ratio equals one in a mean-field
description for short-range pairing interactions because then
$F_{m}= n_{m}=1/2$.  This ratio can deviate significantly from one
when long-range interactions are present, particularly in the
crossover region.  If $F_{m}=n_{m}=1/2$ and the $(+,+)$ and $(+,-)$
edge damping rates are identical, then one can measure $v$ from
their absolute value, since these are proportional to
$(mv^{2}/2\Delta)$ and $\Delta$ follows from $q_{\Delta}$. From the
slope of the scaled damping rate $\gamma_{+,-}(q)$, set out as a
function of the phonon energy $E^{B}_{q}$, one can extract the
square of the coherence length, $- \nabla^{2}_{k}
F_{k}/F_{k}|_{k=k_{m}} = \int d^{3} r F(r) r^{2} \exp(i {\bf k}_{m}
\cdot {\bf r})/ \int d^{3} r F(r) \exp(i {\bf k}_{m} \cdot {\bf
r})$, where $|{\bf k}_{m}| = k_{m}$.  Near crossover, where the
extent of $F$ drops below $k_{m}^{-1}$, this coherence length tends
to the range of $F$.  In the BCS-limit, $k_{m}\rightarrow k_{F}$, $v
\rightarrow v_{F}$ and $- \nabla^{2}_{k} F_{k}/F_{k}|_{k=k_{m}}
\rightarrow (\hbar v_{F}/\Delta)^{2}$ so that the slope tends to
$-1/4$, as shown in Fig.~\ref{fig:scattering_lengths}.  Near the
crossover $q_{\Delta}$ grows and, therefore, it is likely that the
kinematical conditions $|k_{j,+}(E)-k_{j,-}(E')| < q <
k_{j,+}(E)+k_{j,-}(E')$ are violated near the edge. At the minimal
phonon energy $E^{B}_{q}$ for which quasi-particle pairs $j$ are
kinematically excluded, the damping curve exhibits a kink.
We expect the kinks to be smoothed out if the quasi-particles have
finite lifetimes.
In Fig.~\ref{fig:DampingRates}, we show $\Gamma(q)$ for a zero
temperature Cooper-paired system at different detuning values during
an adiabatic sweep through the resonance. The curves were calculated
from a Keyldish-like wavefunction,
$e^{-\alpha\sum_{k}\phi_{k}c^{\dagger}_{{\bf k},+}
c^{\dagger}_{-{\bf k},-}} |0\rangle$, where we choose $\phi_{k}$ as
a Yukawa wavefunction with the range given by the BCS coherence
length, $R=\Delta/\hbar v_{F}$.
While the assumed wavefunction cannot describe the system
quantitatively, the resulting damping rates illustrate the relevant
physics.\\
To demonstrate the practicality of cold atom acoustic attenuation,
we consider a realistic two-component fermion ($^{6}\mathrm{Li}$)-
BEC ($^{23}$Na) mixture ($a_{B}=3.4 \, \mathrm{nm}, \xi=0.14 \, \mu
\mathrm{m}$), at temperature $T=50 \, \mathrm{nK}$, trapped by
harmonic oscillator potentials within a radius $R_{0}=50 \, \mu
\mathrm{m}$ with peak densities $\rho_{F}=10^{12} \,
\mathrm{cm}^{-3}$ and $\rho_{B}=3 \times 10^{14} \,
\mathrm{cm}^{-3}$. We assume that $\overline{a}=5 \, \mathrm{nm}$
and that the system has a gap $\Delta \sim \epsilon_{F}/5$. For
these parameters, $\lambda_{B} \rho_{B} = 266 \, \mathrm{nK}$,
$\epsilon_{F}=610 \, \mathrm{nK}$ ($\Delta = 122 \, \mathrm{nK}$),
the relevant velocities are $v_{F}=4 \, \mathrm{cm/sec}$ and $c=1 \,
\mathrm{cm/sec}$, the time scale for BEC-phonon scattering is
$\Gamma_{0}^{-1} = 0.17 \, \mathrm{msec}$ during which time the
BEC-excitation travels a distance $\Gamma_{0}^{-1} c \sim 1.7 \, \mu
\mathrm{m}$, and the critical phonon momentum for pair breaking,
$q_{\Delta}$, (obtained from $(q_{\Delta} \xi)^2 = -1/2 +
\sqrt{1/4+[\Delta/(\lambda_{B} \rho_{B})]^{2}} \sim 0.178$) is equal
to $q_{\Delta} = 3 \, \mu \mathrm{m}^{-1}$. Creating the wavepacket
by overlapping two laserbeams over a region $L$ of $5 \, \mu
\mathrm{m}$ gives a relative momentum spread $\delta q /q_{\Delta}
\sim 1/(L q_{\Delta}) \sim 0.067$. Measuring the phonon damping or
scattering after a time less than or equal to $0.5 \,
\mathrm{msec}$, this wavepacket would have traveled over less than
$5 \, \mu \mathrm{m}$, so that the measurement can be carried out
within a radius $L_{\rm meas}=5 \, \mu \mathrm{m}$~\cite{Ketterle}.
The relative density variations, only one percent for $L_{\rm
meas}/R_{0}\sim 0.1$ since $\delta \rho_{B}/\rho_{B} \sim (L_{\rm
meas}/R_{0})^{2}$, $\delta \rho_{F}/\rho_{F} \sim (3/2) (L_{\rm
meas}/R_{0})^{2}$, smooth out the sharp features of the damping
curve but only over a few precent of the phonon momentum, less than
the relative momentum uncertainty due to the finite size of the
wavepacket $\sim 5-7 \%$.  Even with a smoothing of ten percent of
the momentum, we expect the most abrupt kinks shown in
Fig.~\ref{fig:DampingRates} to be clearly visible. At finite
temperature, thermal scattering contributes a smooth background that
is considerably smaller than the pair breaking term at sufficiently
low temperatures, such as in the above example. At $\Delta < k_{B}
T$, the thermal occupation of fermion quasi-particles is
exponentially suppressed and so is the corresponding contribution to
the phonon scattering.  The scattering by thermally excited BEC
phonons is not suppressed, but still small: estimating its magnitude
roughly as $\gamma_{\rm therm} \sim 4 \pi \overline{a}^{2} \rho_{B}
c_{B} f$ where $f$ is the fraction of thermally depleted bosons, $f
\sim 0.02$ (from $f \sim (T/T_{\rm BEC})^{3/2}$, where the BEC
critical temperature, $T_{\rm BEC} \sim 20 \, \mu \mathrm{K}$ at
$\rho_{B}= 3 \times 10^{14} cm^{-3}$) gives $\gamma_{\rm therm} \sim
0.03 \Gamma_{0}$.\\
In summary, we have shown that the damping rate of an acoustic
excitation of an overlapping BEC can probe the superfluidity of an
ultra-cold fermionic gas.  The momentum dependence of this rate
yields an unambiguous signature of BCS-like superfluidity, allows a
measurement of the superfluid gap parameter and of the correlation
length in the BEC-BCS crossover regime.  
Kinks in the momentum dependence of the damping rate will reveal
information about the quasi-particle dispersion relations.

\end{document}